\documentclass[%
reprint,
superscriptaddress,
nofootinbib,
nobibnotes,
showkeys,
 amsmath,amssymb,
 aps,
 prl,
floatfix,
]{revtex4-1}
\usepackage{graphicx}
\usepackage{subfigure}
\usepackage{dcolumn}
\usepackage{bm}
\usepackage{amsmath}
\usepackage{float}
\usepackage{lipsum}
\usepackage{color}
\usepackage{times}
\usepackage{textcomp}
\usepackage{latexsym}
\usepackage{mathtools}
\usepackage{acro}
\usepackage[hidelinks]{hyperref}

\newcommand{\pc}{{\textup{pc}}}
\newcommand{\Mlz}{M_{\textup{Lz}}}
\newcommand{\Ml}{M_{\textup{L}}}
\newcommand{\snr}{\rho}
\newcommand{\Mpc}{{\textup{Mpc}}}

\newcommand{\Msun}{{\textup{M}_\odot}}

\newcommand{\Fsis}{F_{\rm SIS}}
\newcommand{\IMBHLimit}{160 \Msun}
\newcommand{\totalDetected}{{\textup{20\%}}}
\newcommand{\pluscross}{{+,\times}}
\newcommand{\mugal}{{\mu_{\textup{gal}}}}

\DeclareAcronym{gw}{
  short = GW ,
  long = gravitational wave ,
  short-plural = s,
}
\DeclareAcronym{bh}{
  short = BH ,
  long = black hole ,
  short-plural = s,
}
\DeclareAcronym{imbh}{
  short = IMBH ,
  long = intermediate-mass black hole ,
  short-plural = s,
}
\DeclareAcronym{smbh}{
  short = SMBH ,
  long = supermassive black hole ,
  short-plural = s ,
}
\DeclareAcronym{dm}{
  short = DM ,
  long = dark matter ,
  short-plural = ,
}
\DeclareAcronym{wimp}{
  short = WIMP ,
  long = Weakly Interacting Massive Particle ,
  short-plural = ,
}
\DeclareAcronym{psd}{
  short = PSD ,
  long = power spectral density ,
  short-plural = ,
}
\DeclareAcronym{snr}{
  short = SNR ,
  long = signal-to-noise ,
  short-plural = ,
}
\DeclareAcronym{et}{
  short = ET ,
  long = Einstein Telescope ,
  short-plural = ,
}
\DeclareAcronym{ligo}{
  short = LIGO ,
  long = Laser Interferometer Gravitational-Wave Observatory ,
  short-plural = ,
}
\DeclareAcronym{lalsuite}{
  short = LALSuite ,
  long = LSC Algorithm Library Suite ,
  short-plural = ,
}
\DeclareAcronym{sis}{
  short = SIS ,
  long = singular isothermal sphere ,
  short-plural = ,
}

\begin{document}

\title{Discovering intermediate-mass black hole lenses through gravitational wave lensing}

\author{Kwun-Hang~Lai}
\email{adrian.k.h.lai@link.cuhk.edu.hk}
\affiliation{Department of Physics, The Chinese University of Hong Kong, Shatin, NT, Hong Kong}
\author{Otto~A.~Hannuksela}
\email{hannuksela@phy.cuhk.edu.hk}
\affiliation{Department of Physics, The Chinese University of Hong Kong, Shatin, NT, Hong Kong}
\author{Antonio~Herrera-Mart\'in}
\affiliation{
SUPA, University of Glasgow, Glasgow, G12 8QQ, United Kingdom
}
\author{Jose~M.~Diego}
\affiliation{Instituto de Física de Cantabria (IFCA, UC-CSIC), Av. de Los Castros s/n, E-39005 Santander, Spain}
\author{Tom~Broadhurst}
\affiliation{Department of Theoretical Physics, University of Basque Country UPV/EHU, Bilbao, Spain}
\affiliation{IKERBASQUE, Basque Foundation for Science, Bilbao, Spain}
\author{Tjonnie~G.~F.~Li}
\affiliation{Department of Physics, The Chinese University of Hong Kong, Shatin, NT, Hong Kong}

\date{\today}
\begin{abstract}
\noindent
Intermediate-mass black holes are the missing link that connects stellar-mass to supermassive black holes and are key to understanding galaxy evolution. 
Gravitational waves, like photons, can be lensed, leading to discernable effects such as diffraction or repeated signals. 
We investigate the detectability of intermediate-mass black hole deflectors in the LIGO-Virgo detector network. 
In particular, we simulate gravitational waves with variable source distributions lensed by an astrophysical population of intermediate-mass black holes, and use standard LIGO tools to infer the properties of these lenses. 
We find detections of intermediate-mass black holes at 98\% confidence level over a wide range of binary and lens parameters.
Therefore, we conclude that intermediate-mass black holes could be detected through lensing of gravitational waves in the LIGO-Virgo detector network. 
\end{abstract}

\maketitle

\section{Introduction}
The existence of stellar-mass and \acp{smbh} has become widely accepted due to X-ray observations of X-ray binary systems~\citep{mcclintock2003black,remillard2006x} and measurements of the
orbits of stars in the center of the 
Milky Way~\citep{ghez2005stellar,ghez2008measuring,kormendy2013coevolution}. 
While the existence of \acp{smbh} is widely accepted, their formation is a mystery due to a \ac{bh} mass gap in the range ($\sim10^2 - 10^5 \Msun$). 
Black holes in this mass range are called \acp{imbh}.
We have yet to observe these \acp{bh} but expect to see a transition from stellar-mass to supermassive \acp{bh}~\citep{sigurdsson1993primordial,ebisuzaki2001missing}. 
Finding this link is crucial to understanding the formation of \acp{smbh} and galaxies. 

Only indirect evidence for \acp{imbh} exists~\citep{mezcua2017observational}, but there are multiple active detection efforts. 
A recent study focusing on mapping the potential of the globular cluster 47 Tucanae through pulsar timing in combination with N-body simulations casts indirect evidence towards an \ac{imbh} in the center of the cluster~\citep{kiziltan2017intermediate}. 
However, the potential for this cluster was derived from N-body simulations subject to a degree of model uncertainty~\citep[see][for a review of the method]{baumgardt2016n}. 
Other forms of searches involve locating X-ray and radio emissions from accretion onto \acp{imbh}, finding tidal disruption events, looking for \ac{imbh} imprints in molecular clouds and microlensing experiments~\citep{van2004intermediate}; for a review, see~\citep{mezcua2017observational}. 
Despite the many efforts to detect \acp{imbh}, the evidence is still inconclusive.

Gravitational lensing is the bending of light, waves or particles near concentrated mass distributions. 
Lensing events probe the \ac{imbh}'s potential, opening a promising avenue for detection. 
On 14 September 2015, the first gravitational wave event was observed with \ac{ligo}~\citep{abbott2016observation}. 
Similarly to light, \acp{gw} can be influenced by gravitational lensing~\citep{ohanian1974focusing,bliokh1975diffraction,bontz1981diffraction,thorne1983theory,deguchi1986diffraction,nakamura1998gravitational,takahashi2003gravitational}. 
When the wavelength of \acp{gw} is comparable to the Schwarzschild radius of the lens, diffraction effects become relevant to the treatment of the lens effect~\citep{nakamura1998gravitational}. 
In the \ac{ligo} band, these wave effects happen in the \ac{imbh} mass range.

There is a growing body of research suggesting that \ac{ligo} will see several lensed gravitational-wave events~\citep{dai2017effect,ng2017precise}. 
\ac{et}, a future ground-based \ac{gw} detector will see a thousand-fold more.
However, previous research has focused on galaxy lensing, while we focus on \ac{imbh} lenses. 
We have calculated the approximate number of \ac{gw} events lensed by \acp{imbh} in \ac{ligo} (\ac{et}), arriving at $\sim 0.05$ ($\sim 50$) events/year. 
For the full calculation, see the "rates" section.

Cao et al. 2014~\citep{cao2014gravitational} investigated the effect of lensing on \ac{gw} parameter estimation using Markov-Chain Monte Carlo to study the lens degeneracy between lens parameters in the \ac{ligo} framework. 
In this work, we show that \acp{imbh} may be detected through lensed events in a realistic LIGO-Virgo detector network.
We use realistic gravitational-wave inspiral merger ringdown waveform~\citep{hannam2014simple} which is utilized in real \ac{ligo} searches. 
By inclusion of spin in our waveform model we account for the possibility that spin precession of the binary~\citep[see][]{blanchet2014gravitational} could mimic lensing. 
In addition, we consider a Advanced \ac{ligo} and Virgo detection network at design sensitivity~\citep{TheLIGOScientific:2014jea,TheVirgo:2014hva}. 
Moreover, we study the parameter constraints in realistic lensing scenarios by including a wide range of lens masses. 

Our results show that lensed \acp{gw} can be used to infer the mass of \acp{imbh}, providing a novel avenue to detect them. 
In particular, if a \ac{gw} is lensed through a potential induced by an \ac{imbh} in our parameter range, we can claim detection with $98\%$ confidence in $\sim \totalDetected$ of the cases. 
Moreover, we show that we can distinguish astrophysical larger than $\sim 8\times 10^3 {\rm AU}$ from \acp{imbh} (typical Schwarzschild radius $\sim 10^{-5} {\rm AU}$). 
Structures smaller than this act effectively as point lenses. 
Finally, we discuss the implications of our results on detection of \acp{imbh}.

\section{Methods} \label{ssec:pmlens}
Consider a system composed of a source emitting \acp{gw}, a lens, and a distant observer.
The source must be close (sub-parsec scale) to the line-of-sight between the lens and the observer for lensing to occur; we denote this distance with $\eta$. 
The angular diameter distances along the line-of-sight between source-lens, source-observer, and lens-observer, are denoted as $D_{LS}$, $D_{S}$, and $D_{L}$, respectively. 
\Acp{imbh} can be approximated as point mass lenses~\citep[][]{takahashi2003gravitational}. 
Given that we ignore the near horizon contribution to the lensing effect, the lensed waveform $h_{\pluscross}^{ \rm lensed}(f)$ is~\citep{nakamura1998gravitational,takahashi2003gravitational}
\begin{equation} \label{eq:amplificationFunctionPointmass}
h_{\pluscross}^{ \rm lensed}(f) = F(w,y) h_{\pluscross}^{ \rm unlensed}(f),
\end{equation}
where
\begin{equation}
\begin{aligned}
F(w,y) ={}& \exp \bigg[ \frac{\pi w}{4} + i\frac{w}{2} \bigg\{ \ln \left( \frac{w}{2} \right) - \frac{(\sqrt{y^{2}+4} - y)^{2}}{4} \\
& + \ln \left( \frac{y + \sqrt{y^{2} + 4}}{2} \right) \bigg\} \bigg] \Gamma \left( 1-\frac{i}{2}w \right) \\
& \times {}_{1}F_{1} \left( \frac{i}{2}w, 1; \frac{i}{2}wy^{2} \right),
\end{aligned}
\end{equation}
where $h_{\pluscross}^{ \rm unlensed}$ is the waveform without lensing, $\Gamma$ is complex gamma function, ${}_{1}F_{1}$ is confluent hypergeometric function of the first kind, $w = 8 \pi \Mlz  f$ is dimensionless frequency, $\Mlz = \Ml (1+z_{L})$ is the redshifted lens mass, $y = D_{L}\eta/\xi_{0}D_{S}$ is the source position, $\xi_{0} = (4\Ml D_{L}D_{LS}/D_{S})^{1/2}$ is a normalization constant (Einstein radius for point mass lens), and $\Ml$ and $z_{L}$ are the lens mass and redshift, respectively. 
The magnification function includes the information of the time delay and is not to be confused with its geometric optics counterpart.
To calculate the magnification function $F(w,y)$, we construct a lookup table, and retrieve its values by bilinear interpolation; 
the error between the table and the exact solution is less than 0.1\%. 
For the \ac{gw} waveform, we use \texttt{IMRPhenomPv2} model, which includes the whole binary inspiral-merger-ringdown phase~\citep{smith2016fast}. 
This assumes an isolated point lens, but we also discuss the effect of external shear and host galaxy in the last section.

We inject \ac{gw} signals from an astrophysical population of binary sources lensed by \acp{imbh} into mock noise data and infer the properties of the \ac{imbh} lens.
The motivation for choosing a distribution of simulated signals is to ensure that we can detect lensed signals across variable lens and binary properties. 
This is in contrast to focusing on a single "example" scenario, which can be fine-tuned. 
Following~\citep[][]{vitale2017impact}, the astrophysical distribution of the binary source is uniform in component masses, dimensionless spin magnitude, and volume; isotropic in spin directions and in sky location.
We assume isolated lenses distributed uniformly in volume, i.e. $P(y) \propto y^{2}$~\citep{takahashi2003gravitational}, where we cut the distribution off at $y>3$ when lensing effects become small ($F\sim 1$) and at $y<0.1$ which makes up only a fraction of the lensed events. 
We distribute redshifted lens mass uniformly in $\Mlz\in [1,1000] \Msun$, which includes the lower \ac{imbh} mass range and extends to stellar-mass range. 
Taking larger masses implies more pronounced lensing effects, and therefore our mass range tests the weak lensing limit. 
If the lens is not isolated, i.e., more lenses are concentrated in the vicinity of galaxies, then the distribution requires corrections. 
These corrections would likely favour nearer sources because most galaxies are at $z\sim 0.3$~\citep{gwyn1996redshift}. 
However, the study of such realistic source distributions requires numerical simulations and is outside the scope of this work. 
Finally, we limit the unlensed \ac{snr} distribution to be $\snr \in [8, 32]$, because \Ac{ligo} requires an \ac{snr} of at least 8 for claiming a detection, and signals with \ac{snr} greater than 32 are rare~\citep{chen2014loudest}. 
We take four different source mass scenarios to investigate the effect of mass ratio on parameter inference. 

\begin{figure}
 \centering
 \includegraphics[width=\linewidth]{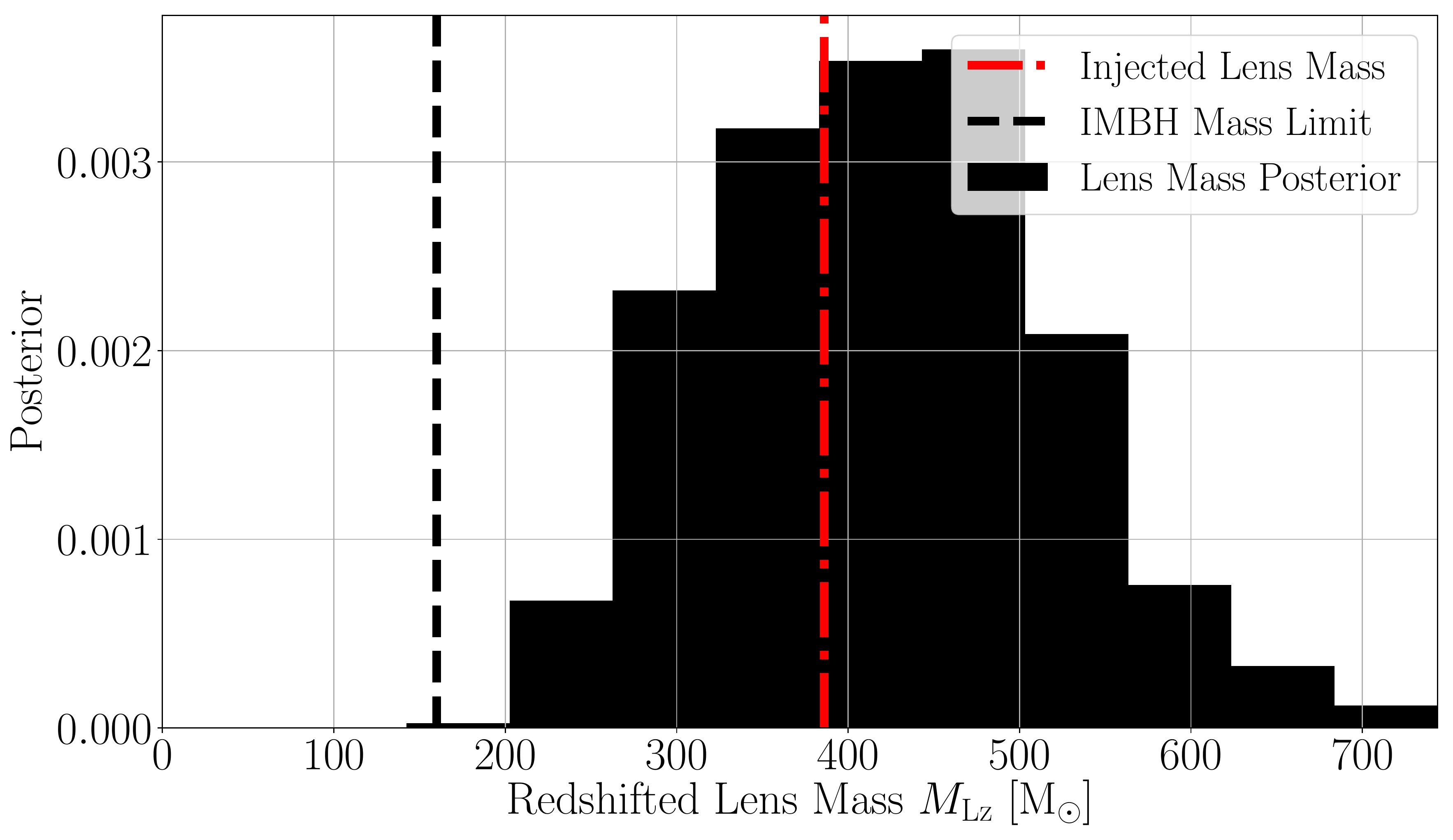}
 \caption{An example redshifted lens mass posterior distribution recovered from an injected, lensed gravitational wave signal using nested sampling (\texttt{LALInference}). 
 The red dashed line shows the injected redshifted lens mass ($\sim 390 \Msun$) and the black dashed line shows the intermediate-mass black hole (IMBH) mass lower bound ($\IMBHLimit$). 
 All of the posterior samples are above the lower bound of IMBH mass.}
\label{fig:posteriorexample}
\end{figure}

We infer the lens mass using nested sampling algorithm  (\texttt{LALInference})~\citep{skilling2004nested}. 
The lens mass and lens redshift are fully degenerate with each other. 
However, in the range detectable by \ac{ligo}~\citep{vitale2017parameter}, the Hubble Deep Field survey shows that the majority of the galaxies that can harbor \acp{imbh} are located at $z_L \sim 0.6$~\citep{gwyn1996redshift}, which can be used as an approximate, typical redshift in our analysis.
Therefore, we choose the probability $P(\Mlz>\IMBHLimit) > \rm 98\%$ to indicate a successful detection of \ac{imbh}. 
We show an example redshifted lens mass posterior distribution recovered from an injected \ac{gw} that passes through a lens of mass $\Ml \approx 380 \Msun$ (Fig.~\ref{fig:posteriorexample}). 
The posterior peaks around the injected value and the samples are above the \ac{imbh} mass limit. 
In our analysis, this posterior is classified as detection.

\section{Lensing Event Rates}
The number of \ac{gw} events lensed by \acp{imbh} may be estimated using the known \ac{gw} event rates and assuming an \acp{imbh} lens population.
Astrophysical modeling suggests that around 20\% of globular clusters could harbor \acp{imbh}~\citep{coleman2002production}. 
In that case, by order of magnitude, we have $\sim 10^2-10^4$ \acp{imbh} per galaxy~\citep{fall2001dynamical,kruijssen2009interpretation}, which is in agreement with N-body simulations of molecular clouds reach similar number of \acp{imbh} in galaxies~\citep{shinkai2017gravitational}.
We assume a typical $n\sim 0.03\,\rm Mpc^{-3}$ density of galaxy lenses at $z_L\sim 0.3-0.6$~\citep{cole20012df,bell2003optical}, angular diameter distance to lens $D_L\sim 800$ Mpc, and from lens to source $D_{LS}\sim 800$ Mpc. The probability of a single event being lensed is given by the area of the lens in the lens plane, which to the first order can be computed as the area within the Einstein radius, divided by the total area of the lens plane. The Einstein radius of the \ac{imbh} lens within a galaxy is boosted by a typical galaxy magnification $\mu\sim 2-3$, which also boost the probability of it being lensed.
The rate of unlensed events is $\sim 800-10000$ events/year at design sensitivity of Advanced \ac{ligo}~\citep{scientific2017gw170104,ng2017precise} (and $1000$ times more at \ac{et} sensitivity), based on rates inferred directly by LIGO.
Therefore, the total number of lensed events boosted by magnification $\mu$ is $R_{\rm lensed}\sim 10^{-14} (\Ml / M_\odot) N_{\rm IMBH} N_{\rm GW} \mu^{5/2} \sim 3.74\times10^{-6}-0.16$~$\text{events}/\text{year}$ at design sensitivity ($\sim 3.74\times10^{-3}-160$~events/year in \ac{et}), where the lower and upper bound are given by pessimistic and optimistic parameters respectively.
However, taking typical \ac{imbh} candidate mass~\citep{coleman2002production,lou2012intermediate} and lens populations~\citep{caputo2017estimating} as an example yields $R_{\rm lensed}\sim 10^{-14} (\Ml/M_\odot) N_{\rm IMBH} N_{\rm GW} \mu^{5/2} \sim 10^{-14}\times (5000 M_\odot/M_\odot)\times 10^4\times 6000\times 3^{5/2} \sim0.05$~events/year (50 events/year in \ac{et}). 
This is roughly comparable to the microlensing event rates for \acp{imbh} in the electromagnetic band, which stand at $0.86$~events/(20 years)~\citep{kains2016searching}.
The $5000 M_\odot$ is a more massive lens than what we consider as a reference in our nested sampling study, but our results are applicable in this mass regime as well because it is easier to detect larger lens masses tend due to larger lens effects.
Such a number assumes that \ac{imbh} candidates are within the range of thousands of solar masses~\citep{coleman2002production,lou2012intermediate} and the number of \acp{imbh} is around $10^4$ per galaxy~\citep{caputo2017estimating}.

We stress that having precise event rate estimates is difficult due to the large uncertainty in the number density of \acp{imbh}, the event rates of binary coalescences, uncertainty in \ac{imbh} mass distribution and the uncertainty in the lens magnification distribution.
Therefore, the event rate should be taken as an order-of-magnitude estimate demonstrating that detecting lensing by \acp{imbh} is possible.
Nevertheless, if we do detect a \ac{gw} signal lensed by an \ac{imbh}, we have a chance to discriminate it using the methods we outline in this work.

\section{Detecting intermediate-mass black holes} \label{ssec:lensparamconstraint}
\begin{figure}
 \includegraphics[width=\linewidth]{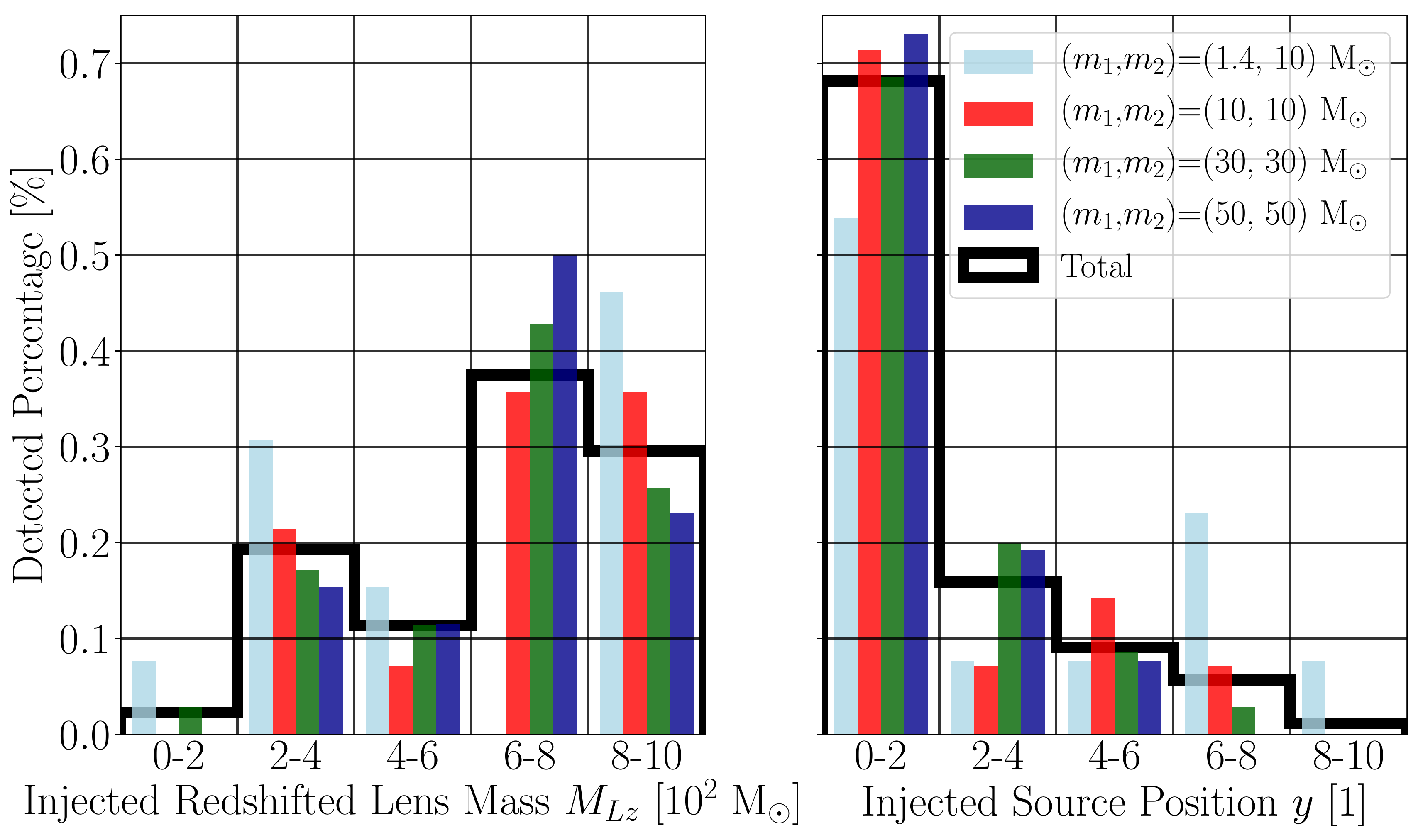}
 \caption{Detected intermediate-mass \acp{bh} as a function of injected redshifted lens mass $\Mlz$ (left panel), and source position squared $y^2$ (right panel) for four different source binary masses and their sum.
Detection is defined at $98\%$ confidence level. 
 The number of detections decreases with increasing source positions, and increases with increasing lens masses.
}
  \label{fig:IMBHDetectionLensMassAndPosition}
\end{figure}

We find detections over a wide range of lens masses ($\Mlz \gtrsim 200 \Msun$), and find a rising trend in detections with higher lens mass (Fig.~\ref{fig:IMBHDetectionLensMassAndPosition}, left panel). 
Of these, we find around $\sim 16-30$\% of detected \acp{imbh} with relatively small redshifted lens masses ($\Mlz<500$; Fig.~\ref{fig:IMBHDetectionLensMassAndPosition}). 
Approximately $\totalDetected$ of lenses are detectable in our parameter range.
However, there are two false alarms with masses lower than $\IMBHLimit$, which is statistically expected at 98 \% confidence level, given that we have over 100 detections.

In addition to redshifted lens mass, we characterize the effect of source position on the detectability of \acp{imbh}. 
The source position $y$ is proportional to the horizontal distance from the line-of-sight. 
Because smaller source positions $y$ correspond to larger lens effects, we expect better constraints at small $y$. 
Indeed, we detect a more substantial number of \acp{imbh} at low source positions, where more than 55\% of them are in the range $y^2~=~[0, 2.5]$ for all source masses (Fig.~\ref{fig:IMBHDetectionLensMassAndPosition}, right panel). 
Meanwhile, we find that there are also detections at relatively large source positions ($y^2>5$) but the number decreases for increasing position.
The source position at $y~=~\sqrt{2.5}\approx1.58$ can be translated back to the displacement from the line-of-sight. 
Assuming typical lens-to-source distance $D_{LS}=300\Mpc$, lens distance $D_L=300\Mpc$ and source distance $D_S=600\Mpc$, we have $\eta\approx 0.01 \pc \sqrt{\Ml/ \Msun}$. 
Hence, the line-of-sight distance where we detect \acp{imbh} is likely sub-parsec. 

We detect \acp{imbh} across the \ac{snr} range $\snr \in [9,32]$. 
To put this into the context of the current \ac{ligo} detections, all of the confirmed detections have had a network inferred \ac{snr} inside our range (see the first observing run summary \citep{abbott2016binary}).

In all four classes of source mass realizations, we detect around $\totalDetected$ of the \acp{imbh} at a 98\% confidence level. 
Among the detected signals, we also compare the Bayes factors between the lensed model and the unlensed model. 
The evidence for the lensed hypothesis is significantly ($400$ times) larger than the unlensed hypothesis for more than 70\% of the signals, which suggests that these detections are not cause by noise. 
Moreover, we have simulated and analyzed a set of unlensed signals. 
Of these, none prefer the lensed hypothesis at Bayes factor above 40. 
The lensing effect is not degenerate with sky location and other parameters, and therefore calibration uncertainties are not expected to affect the results drastically~\citep[see e.g.][for review]{tgflivitale2001,cahillane2017calibration}. 
Therefore, we are confident that the detection criteria $P(\Mlz>\IMBHLimit) > \rm 98\%$ together with the Bayes factor analysis provide a reasonable estimate of the detectability of \ac{imbh}. 
We have also analyzed the first \ac{gw} event GW150914~\citep{abbott2016observation}, finding no evidence of lensing (Bayes factors both being the same up to 4th significant digit for the lensed and unlensed case).

In conclusion, we find detections across $\Mlz \in [160, 1000] \Msun$, $y^2\in[0.01,9]$ and $\snr \in [9,32]$, and find that higher lens masses, smaller source positions and higher \acp{snr} are favored. 

\section{Discriminating between point and finite-size lens} \label{ssec:match}
\noindent
Other small astrophysical lens objects could mimic \ac{imbh} lenses. 
We study a finite-size \ac{sis} model to test our ability to discriminate between finite and point lenses using \acp{gw}.
If the size is small enough, the object will collapse into a \ac{bh}.
The \ac{sis} model represents the approximate mass distribution of an extended astrophysical object. 
Its magnification function~\citep{takahashi2003gravitational}
\begin{equation} \label{eq:sis}
\begin{aligned}
\Fsis(w,y) ={}& -iwe^{iwy^{2}/2} \int_{0}^{\infty} dx \; \bigg\{ x J_{0}(wxy) \\
& \times \exp \bigg[ iw \bigg( \frac{1}{2}x^{2}-x+y+\frac{1}{2} \bigg) \bigg] \bigg\},
\end{aligned}
\end{equation}
where $w=8\pi \Mlz f$, $\Mlz = 4\pi^{2} v^{4} (1+z_{L}) D_{L} D_{LS}/D_{S}$ is the redshifted mass inside the Einstein radius $\xi_0$, $v$ is a characteristic dispersion velocity of the model and $x=|\vec{\xi}/\xi|$ is the normalized impact parameter. 
We expect the \ac{sis} model to be indistinguishable from an \ac{imbh} model due to mass screening effect when the Einstein radius $\xi_0$ is small. 

In order to compare the \ac{sis} and the point lens model, we compute the match $\mathrm{m}(h_{a},h_{b})$~\citep{Canton:2014ena,Usman:2015kfa} between two waveforms $h_{a}$ and $h_{b}$ maximized over time, phase and amplitude.
For this comparison, we simulate \acp{gw} from a (30,30) $\Msun$ source oriented in the overhead direction and compare the match between the signals lensed by an \ac{sis} and a point lens.

\begin{figure}
 \includegraphics[width=\linewidth]{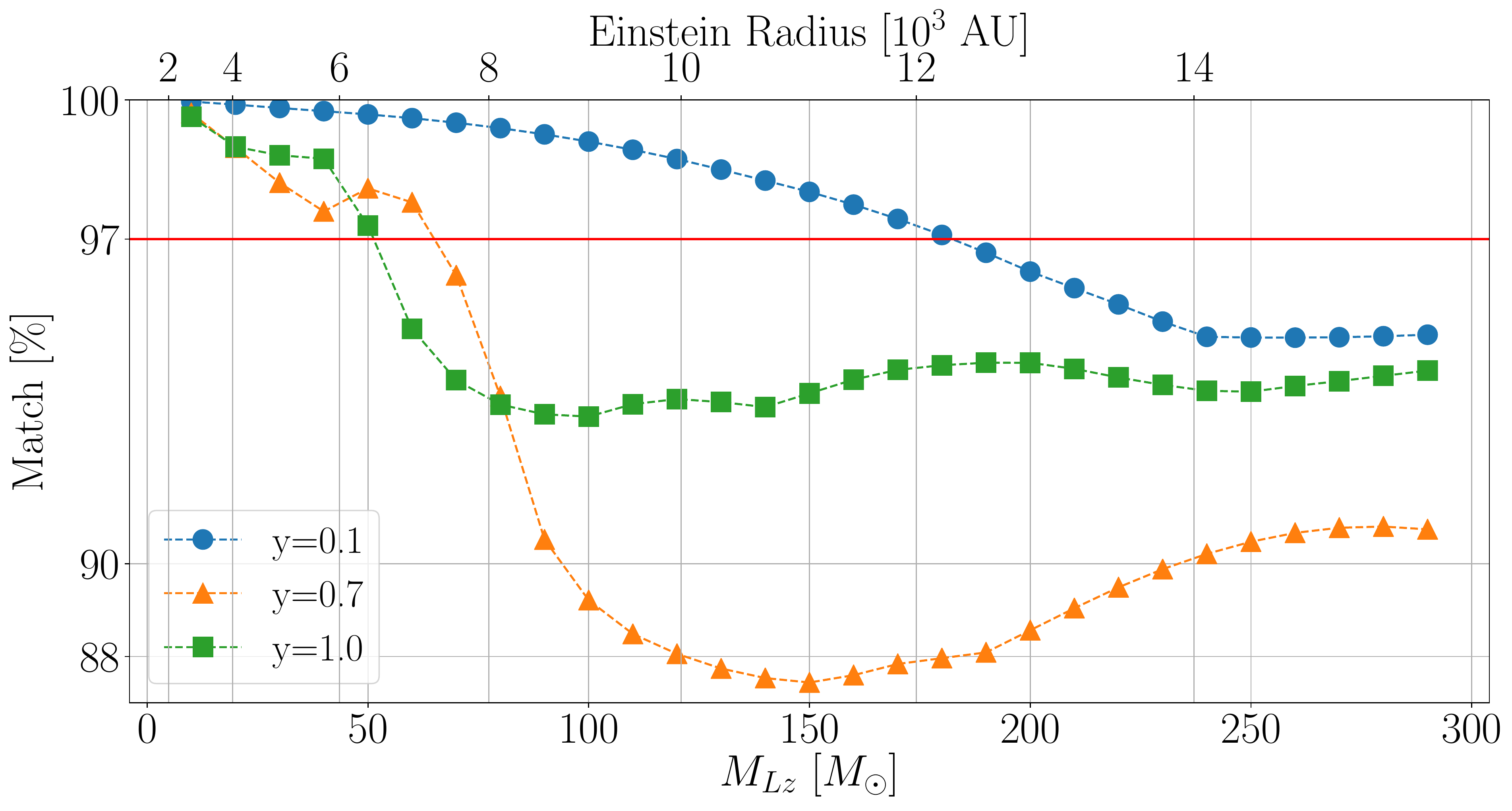}
 \caption{
Match $\mathrm{m}(h_{a},h_{b})$ between waveform lensed by an SIS and a point mass lens maximized by non-linear least-squares fitting as a function of redshifted lens mass $\Mlz$ (bottom axis) and Einstein radius (top axis). 
Three source positions $y=0.1$, $y=0.7$ and $y=1.0$ are shown as dashed lines with blue circles, orange triangles, and green squares, respectively. 
The red horizontal line denotes $97\%$ match. 
The source and source-to-lens angular diameter distances are chosen so that the lens is in the middle with $D_{\rm L} = D_{\rm LS} = 400 \Mpc$. 
At redshifted lens mass $\Mlz=200 \Msun$ all matches are below $97\%$. 
}
 \label{fig:sismatch}
\end{figure}

We consider different pairs of $(\Mlz,y)$, and maximize the $\mathrm{m}(h_{a},h_{b})$ by non-linear least squares fitting and classify $\mathrm{m}(h_{a},h_{b}) < 97\%$ as distinguishable in \ac{ligo} waveform following~\cite{hinder2017eccentric}. 
We show that the \ac{sis} lens and point lens can be discriminated when redshifted lens mass $\Mlz > 200 \Msun$, shown as a match lower than 97\% in Fig.~\ref{fig:sismatch}. 
The source positions $y=1$  and $y=0.1$ show higher match for all redshifted lens masses because small source positions $y$ cause only a total magnification of the signal, while very large $y$ cause only small lens effect. The oscillatory property of the magnification functions \textit{F} and $\Fsis$ induces the oscillatory dependency between the match and $\Mlz$.

The SIS model has an intrinsic length scale, which is the Einstein radius.
Since astrophysical structures with diameters smaller than $10^{4} \rm AU$ show high match (Fig.~\ref{fig:sismatch}), they can not be discriminated from point lenses. 
Indeed, our results suggest we can distinguish an \ac{imbh} from a globular cluster (half-mass radius at pc scale~\citep{van2010comparison}), but not structures smaller than $10^4$~AU. 

\section{Discussion and conclusions} \label{sec:conclusions}
We demonstrate that it is possible to discover \acp{imbh} in the LIGO-Virgo network by analyzing \acp{gw} lensed by these \acp{bh} even for relatively small lens masses ($\Ml\sim200-300\rm \Msun$). 
We find that in $\sim \totalDetected$ of cases the effect of lensing is strong enough to discover an \ac{imbh} with 98\% confidence in our parameter range. 
Moreover, we find that we can discriminate between \ac{sis} and point lens models when the Einstein radius of the \ac{sis} is larger than $10^{4} \rm AU$. 
In particular, our results suggest that we may discriminate an \ac{imbh} lens from an extended astrophysical object, but it is hard to distinguish between \ac{imbh} lenses and compact objects of similar mass. 
However, there is currently no conclusive evidence of compact objects with masses greater than $200$ $\Msun$. 

In our results, we do not account for shear effects by host galaxies. 
However, it is important to discuss its effect on the results, as compact objects are typically discovered as part of a galaxy.
Such shear magnifies the \ac{gw} signal and introduces a degeneracy between the inferred lens mass and shear magnification.
In particular, external shear enlarges the point lens' Einstein radius, stretching it along the deflection field of the host galaxy and changing the lens time delay~\citep[see][]{diego2017dark}.
Consequently, the effective mass of the lens becomes $\Ml^\prime \rightarrow \mu_t \times \Ml$ owing to its dependence on the Einstein radius.
The stretching is modest when the magnification by galaxy $\mugal$ is reasonably low ($\mugal \lesssim5$), and the new radius is larger by a factor of $\sim \mu_t^{1/2}$, with $\mu_t$ being the tangential magnification component.
The lensing probability at high magnification goes as $\mugal^{-2}$.
As a consequence, typical magnifications are modest, between $\mugal\sim 1-3$. 
Taking such typical shear and magnification, we would need to measure 300 $\Msun$ lens to distinguish the lens as an \ac{imbh}. 
Meanwhile, the magnification in shear would boost the \ac{gw} event rates.

In contrast with previous results, our results imply that we can detect \acp{imbh} within \ac{ligo} data.
However, there is also an interesting prospect of detecting stellar mass \acp{bh} with \acp{gw}. 
\ac{ligo} may not be sensitive enough to constrain the properties of $\sim 1 \Msun$ lenses and the event rate required for \acp{gw} lensed by $\sim 30 \Msun$ lenses with high enough \ac{snr} may be too low, but there is an interesting prospect of detecting these \acp{bh} with future third-generation detectors such as the Telescope and Cosmic Explorer~\citep[see][] {3gdetectors1,3gdetectors2,3gdetectors3,3gdetectors4}; these prospects are discussed by~\citep{christianetalcompanion}.

Moreover, \ac{imbh} could be directly detected by \ac{ligo}; however, these detections are limited to a mass range $M\lesssim 150 \rm M_\odot$ due to the low-frequency noise in \ac{ligo}~\citep{shinkai2017gravitational}. 
Our method does not suffer from such cut-off, and its discriminatory power increases for more massive lenses.

In conclusion, we have shown that lensing of \acp{gw} by \acp{imbh} is detectable over a wide range of parameters and that a detection of a point mass lens of mass higher than $300 \Msun$ in principle warrants a discovery of \acp{imbh}. 
In the future, we will expand our study on the effect of different lensing models, and mixed models with \acp{bh} and surrounding matter; 
for example, it is essential to investigate lens models with globular clusters containing \acp{imbh} and lenses admixed in shear.

\bibliographystyle{unsrt}

\end{document}